\documentclass[10pt,a4paper]{article}
\usepackage{graphicx}
\usepackage{subfigure}
\usepackage{amsmath,amsthm}
\usepackage{amssymb,amsfonts}
\usepackage{authblk}
\usepackage{url}
\usepackage{hyperref}
\usepackage{booktabs}
\usepackage{multirow}

\newtheorem{definition}{\textbf{Definition}}

\title{Copula Entropy based Variable Selection for Survival Analysis}
\author{Jian MA\thanks{Email: majian@hitachi.cn}}
\affil{Hitachi China Research Laboratory}
\date{}

\begin{document}

\maketitle

\begin{abstract}
	\noindent
	Variable selection is an important problem in statistics and machine learning. Copula Entropy (CE) is a mathematical concept for measuring statistical independence and has been applied to variable selection recently. In this paper we propose to apply the CE-based method for variable selection to survival analysis. The idea is to measure the correlation between variables and time-to-event with CE and then select variables according to their CE value. Experiments on simulated data and two real cancer data were conducted to compare the proposed method with two related methods: random survival forest and Lasso-Cox. Experimental results showed that the proposed method can select the 'right' variables out that are more interpretable and lead to better prediction performance.

\end{abstract}

\section{Introduction}
Variable Selection (VS) is an important problem in statistics and machine learning. It arises when there are many candidate variables for building model while only a subset of them is related to the output of the model. With selected variables as input, the built model will be interpretable to end-users. This is of critical importance in the fields that require interpretability, such as medicine and law. In high dimensional cases, VS can also improve the efficiency of the model with fewer variables. 

Copula Entropy (CE) is a mathematical concept defined for measuring statistical independence \cite{Ma2011}. It is based on copula theory \cite{nelsen2007,joe2014} and actually a type of Shannon entropy. Ma and Sun proved the equivalence between CE and mutual information in information theory \cite{Ma2011}. CE is an ideal correlation/independence measure with many good properties, such as multivariate, non-positive, invariance to monotonical transformation, equivalence to correlation coefficient in Gaussian cases. It has be applied to solve VS problem in \cite{Ma2021} by measuring the correlation between variables and target variable. The advantage of the CE-based method over traditional VS methods has also been shown with real data. CE-based VS method has been widely applied to different fields, such as hydrology \cite{Chen2013}, medicine \cite{Mesiar2021}, manufacturing \cite{Sun2021}, reliability \cite{Sun2019}, energy \cite{Liu2022}, etc.

Survival analysis is a special type of regression problem that try to predict time-to-event, i.e., time till an event occurs, from a group of variables. The time-to-event data is special also because of censoring mechanism for the case that event does not occur during observation. Survival analysis has wide applications in the fields, such as medicine, manufacturing, and social sciences. There are many existing methods for survival analysis, such as parametric Cox model \cite{Cox1972}, non-parametric Random Survival Forest (RSF) \cite{Ishwaran2008}, and many others. See \cite{Wang2019} for a survey on survival analysis methods.

VS is also important for the application of survival analysis where interpretability is needed. For example, when building a prognostic model for cancer survival, clinicians should understand why the model works for his/her clinical decisions. It requires the method can select the meaningful variables for prognostic models. 

There are several VS methods for survival analysis. Based on parametric Cox model, Lasso-Cox was proposed by adding a penalized term to the likelihood of Cox models \cite{Yang2013}. RSF can also select important variables out based on the observation that important variables tend to be close to the root node of survival trees \cite{Ishwaran2010}.

In this paper we propose to apply CE to VS problem for survival analysis. Similar to \cite{Ma2021}, CE will be used to measure the correlation between variables and target variable (time-to-event) and then one can select important variables according to CE value. Since CE is non-positive, the variables with smaller CE will be preferred.

This paper is organized as follows: Section \ref{sec:related} introduce the related works, The theory and estimation method of CE will be introduce in Section \ref{sec:ce}, Section \ref{sec:method} presents the proposed method, experiments with simulated and real data will be presented in Section \ref{sec:exp}, Section \ref{sec:discussion} is some discussion, we conclude the paper in Section \ref{sec:con}.

\section{Related works}
\label{sec:related}
\subsection{Regularized-Cox}
Regularization is a typical methodology for variable selection. By adding a penalized term to likelihood-based objective function, it tends to squeeze the size of covariates with non-zero coefficients in the model. It can be applied to any parametric regression models with known probability density function, including Cox's proportional hazard model. Yang and Zou \cite{Yang2013} introduced a cocktail algorithm for solving the penalized Cox model that enjoys a good convergence property. The penalized term of the algorithm is defined with L1 and L2 norm, as follows:
\begin{equation}
	P(\alpha) = \alpha|\beta|_1 + \frac{1}{2}(1-\alpha)|\beta|_2,
\end{equation}
where $|\cdot|_n$ denotes $L_n$ norm and $0<\alpha\leq1$. With $\alpha =1$, it is for Lasso-Cox, a special case of Lasso \cite{Tibshirani1996}.

\subsection{Random Survival Forest}
Random Survival Forest \cite{Ishwaran2008} is a recently introduced method for survival analysis and variable selection. It is an ensemble-based method that build a group of survival tree on the bootstrap samples from time-to-event data. The RSF algorithm consists of four steps: 1) generating a group of bootstrap samples from training data; 2) growing a survival tree on each sample; 3) calculating the cumulative hazard function for each tree; 4) making prediction for testing data by averaging over all the output of the above trees. As traditional random forest method, RSF is a non-parametric method without any model assumptions. Please refer to \cite{Wang2017} for a review on RSF.

A RSF based method \cite{Ishwaran2010} has been developed for variable selection. It is based on the observation that in RSF, important covariates tend to be close to the root node of survival trees. Therefore, they can be selected by the so-called minimal depth method.

\subsection{Copula-based survival analysis}
Copula has been applied to survival analysis recently. Petti et al \cite{Petti2022} proposed a copula-based model for bivaiate survival data with various censoring mechanisms. Vine copula, an approach to built multivariate copula, was proposed to built models for time-to-event prediction \cite{Pan2022}. Bentoumi et al \cite{Bentoumi2019} proposed a parametric copula based dependence measure for measuring the dependence between survival time and covariates for length-biased survival data. Copula was also used to model the dependence between event times \cite{Emura2018}. Please see \cite{Govindarajulu2020} for more recent applications of copula to survival analysis.

\section{Copula Entropy}
\label{sec:ce}
Copula theory is a probabilistic theory on representation of multivariate dependence \cite{nelsen2007,joe2014}. According to Sklar's theorem \cite{sklar1959}, any multivariate density function can be represented as a product of its marginals and copula density function (cdf) which represents dependence structure among random variables. 

With copula theory, Ma and Sun \cite{Ma2011} defined a new mathematical concept, named Copula Entropy, as follows:
\begin{definition}[Copula Entropy]
	Let $\mathbf{X}$ be random variables with marginals $\mathbf{u}$ and copula density function $c$. The CE of $\mathbf{X}$ is defined as
	\begin{equation}
	H_c(\mathbf{x})=-\int_{\mathbf{u}}{c(\mathbf{u})\log c(\mathbf{u})d\mathbf{u}}.
	\label{eq:ce}
	\end{equation}	
\end{definition}

A non-parametric estimator of CE was also proposed in \cite{Ma2011}, which composed of two simple steps:
\begin{enumerate}
	\item estimating empirical copula density function;
	\item estimating the entropy of the estimated empirical copula density.
\end{enumerate}
The empirical copula density in the first step can be easily derived with rank statistic. With the estimated empirical copula density, the second step is essentially a problem of entropy estimation, which can be tackled with the KSG estimation method \cite{Kraskov2004}. In this way, a non-parametric method for estimating CE was proposed \cite{Ma2011}.

\section{Proposed Method}
\label{sec:method}
CE has been applied to variable selection problem \cite{Ma2021} by measuring the correlation between covariates and response with CE. In this section we propose to use this method for variable selection in survival analysis. The idea is to measure the correlation between variables and time-to-event with CE and then selection the variables that mostly associate with survival time. Since CE is non-positive, the variables associated with smaller CE will be selected. The smaller CE value, The better the variable. CE is estimated with the nonparametric method in \cite{Ma2011} and therefore our method is model-free. Another advantage of our method is tuning-free since the used method for estimating CE is insensitive to hyperparameters.

\section{Experiments}
\label{sec:exp}
\subsection{General setting}
We test our method on both simulated data and two real data. We also compare it with two important related methods: RSF and Lasso-Cox. In simulation experiment, we will simulate a sample of survival model with known variables and compare the ability for variable selection of these three methods with reference to the model setting. In real data experiments, we compare these three methods from two aspects: interpretability and predictability. For interpretability, we first use these three methods to select a group of variables for each data and then check the interpretability of the selected variables with reference to domain knowledge. For predictability, we will fit a survival regression model with the selected variables with our method. As contrast, we will fit another survival regression model with all the variables in the dataset. RSF and Lasso-Cox will also be applied to the data. And then the prediction results on the data will be compared in terms of two common performance measures: Mean Average Error (MAE) and C-Index. 

Five \textsf{R} \cite{RCT2022} packages were used for the data and the implementations of the methods in the experiments. The \texttt{survsim} package \cite{Morina2014} was used for simulating survival data. The real data in the experiments is the `cancer' and `veteran' data in the \texttt{survival} package. The \texttt{copent} package \cite{ma2021copent} was used for the method for estimating CE. the \texttt{survreg} function in the \texttt{survival} package was used for fitting survival regression models. The \texttt{randomForestSRC} \cite{Ishwaran2007,Ishwaran2008,Ishwaran2022} and \texttt{fastcox} \cite{Yang2013} packages were used for the implementation of RSF and Lasso-Cox. The default parameter values were used in the experiments for all the functions of the methods in these packages.

\subsection{Simulation}
\label{sec:sim}
To compare our method with RSF and Lasso-Cox, we simulated a right-censored survival data from 1000 subjects with a maximum follow-up time of 100 days and 5 covariates. We assumed individual homogeneity in simulation. The time-to-event follows Weibull distribution with two parameters (ancillary parameter $a_0$ and $\beta_0$). The 5 covariates all follow normal distribution in the experiments with different mean and variance. The time-to-censoring distribution is also Weibull distribution with two parameters (ancillary parameter $a_c$ and $\beta_c$). If time-to-event is smaller than time-to-censoring, then the sample is considered to be observed, otherwise censored. In the experiment the parameters of the simulation were set as listed in Table \ref{tab:sim}.

\begin{table}
	\centering
	\caption{The paramters in the simulation experiments.}
	\begin{tabular}{l|c|c|c}
		\toprule
		&Distribution&Parameters&Value\\
		\midrule
		\multirow{2}{*}{time-to-event}&\multirow{2}{*}{Weibull}&$a_0$&2\\
		\cline{3-4}
		&&$\beta_0$&1\\
		\hline
		\multirow{2}{*}{time-to-censoring}&\multirow{2}{*}{Weibull}&$a_c$&0.85\\
		\cline{3-4}
		&&$\beta_c$&5\\
		\hline
		\multirow{10}{*}{covariates}&\multirow{5}{*}{/}&\multirow{5}{*}{$\beta$}&1.4\\
		&&&1.2\\
		&&&0\\
		&&&1.2\\
		&&&0.2\\
		\cline{2-4}
		&\multirow{5}{*}{normal}&\multirow{5}{*}{($\mu$,$\delta^2$)}&(0.4,1.1)\\
		&&&(1.0,1.1)\\
		&&&(0.7,1.1)\\
		&&&(0.2,1.3)\\
		&&&(0.2,1.1)\\
		\bottomrule
	\end{tabular}
	\label{tab:sim}
\end{table}
In the simulation, the coefficients of the covariates were set with different values to test the effectiveness of the three methods. The covariates with larger coefficients value are expected to be selected. The covariates with same coefficients value are supposed to be treated as equally important. 

\begin{figure}
	\centering
	\includegraphics[width=0.9\linewidth]{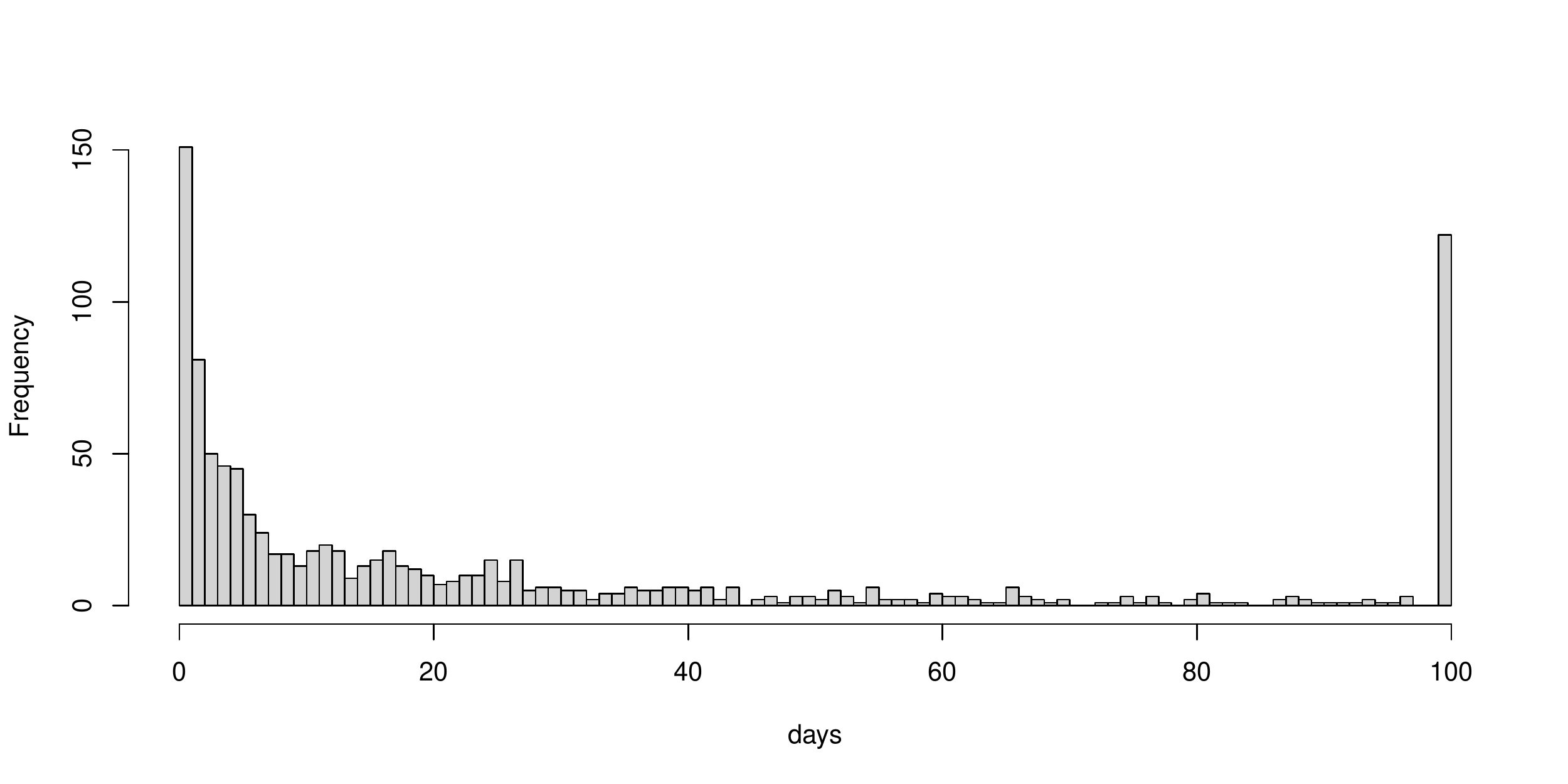}
	\caption{Histogram of the time-to-event of the simulated data.}
	\label{fig:hist}
\end{figure}

The distribution of time-to-event of the simulated data is illustrated in Figure \ref{fig:hist}, from which it can be learned that a part of data is right-censored at the end of the follow-up time.

\begin{figure}
	\centering
	\includegraphics[width=0.95\linewidth]{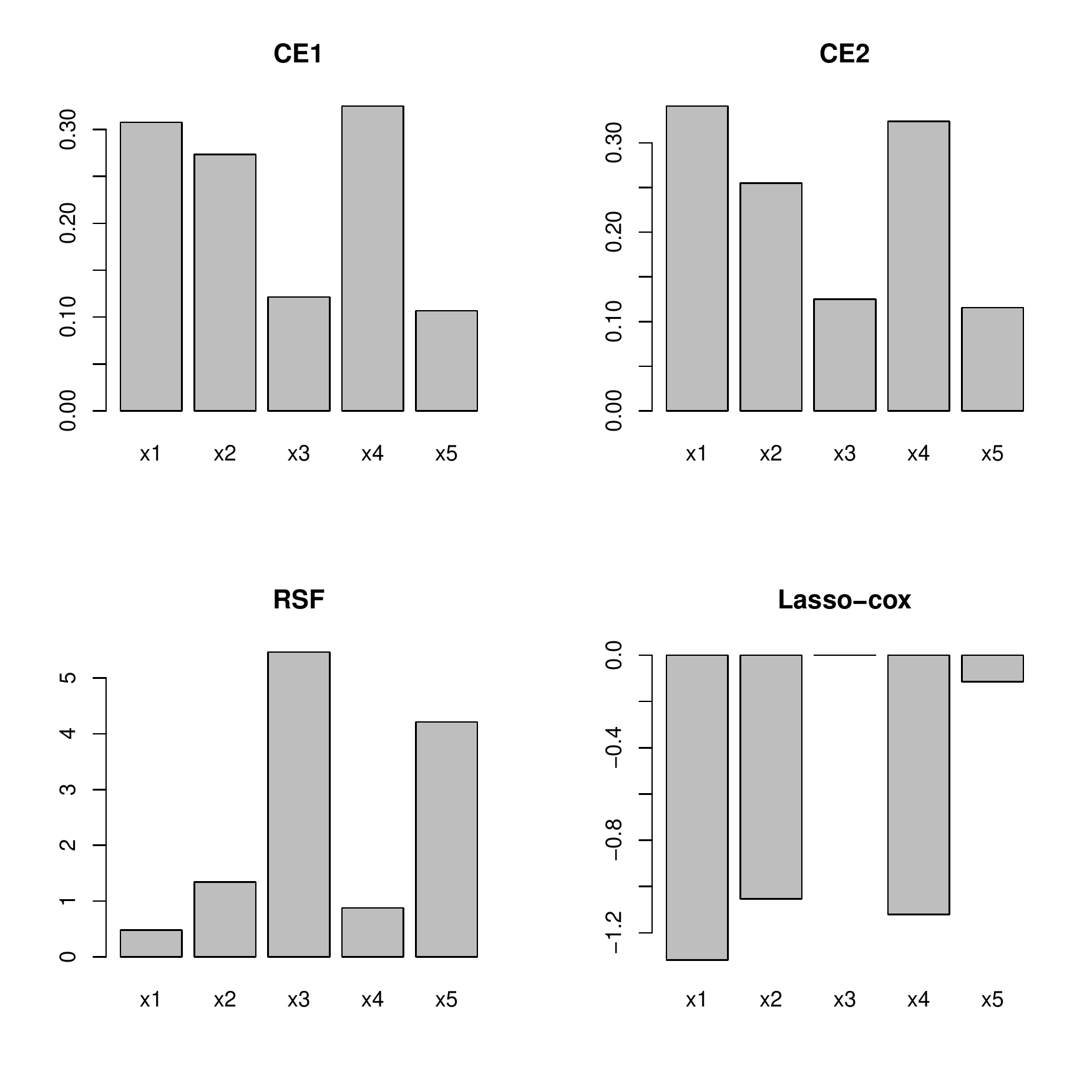}
	\caption{The variable selection results of the three methods in the simulation experiment. CE1 and CE2 are the results of our method without and with censor information respectively.}
	\label{fig:sim1}
\end{figure}

We then applied our method, RSF, and Lasso-Cox to the simulated data. For our method, two results were derived: one from estimating CE between time-to-event and covariates and the other from estimating CE between time-to-event, censoring mark, and covariates. This is for testing whether censoring information can improve the performance of our methods. For RSF, variables are selected using the minimal depth method. For Lasso-Cox, we have two hyperparameter to choose: $nlambda$ and $\alpha$. In the experiment, we set $nlambda=20$ and $\alpha=1$ to estimate a group of $\beta$ with Lasso-Cox model with L1 norm regularization. Then we choose a $\beta$ that fit the coefficients of the simulated model. The variable selection result of these three methods are shown in Figure \ref{fig:sim1}. It can be learned that all the methods present the right variable importance to the covariates and that adding censored information cannot improve the performance of our method too much. So in real data experiment, we will use our method without censor information only.

\subsection{Real data}
Two real data in the \texttt{survival} package in \textsf{R} were used to compare our method with RSF and Lasso-Cox. The first data is the cancer data collected from the patients with advanced lung cancer from the US North Central Cancer Treatment Group. It contains 228 samples with 10 variables, including survival time, censoring status, age, sex, ECOG performance score and Karnofsky performance score by physicians, Karnofsky performance score by patients themself, calories consumed at meals, and weight loss in last six months. After omitting the samples with missing values, the remaining 167 samples were used in the experiment. The second data is the veteran data collected from 137 lung cancer patients from the US Veterans Administration, of whom 69 patients received the standard test, 68 patients received Chemotherapy test. Finally, 64 patients died for each test and only 8 patients kept alive. The dataset contains 9 variables, including test type, cell type of tumors, survival time, censoring status, Karnofsky performance score, months from diagnosis to randomization, age, and the indicator for whether received treatment before the recent one. All the 137 samples were used in the experiment.

From each dataset, four models for survival analysis were built. We first fit a survival regression model to the data with all the variable included. Second, we estimated CE between survival time and other variables and then selected the variables with large CE value as the input to another survival regression model built with the data with only selected variables. Third, we fit a RSF model to each data and check the variables selected by RSF. Last, we fit a Lasso-Cox model to each data and choose a variable set with similar size as other methods by tuning the hyperparameter of Lasso-Cox. We compared the variables selected by these four models from the viewpoint of interpretability. To further compare the models, the prediction results of these models were derived from the whole dataset. The prediction performance is measured with two measures: MAE and C-index.

The variable importance of the four models is defined with different ways. The coefficient of the covariates of the survival regression model are considered as variable importance. For our method, CE between covariates and time-to-event measures the importance of variables. In RSF, variables are selected with tree minimal depth methodology \cite{Ishwaran2010} and a measure of variable importance are computed simultaneously. For Lasso-Cox model, a group of variables are selected by choosing a $\beta$ so that the number of the selected variables are similar to that of our methods.

\begin{figure}
	\centering
	\includegraphics[width=\linewidth]{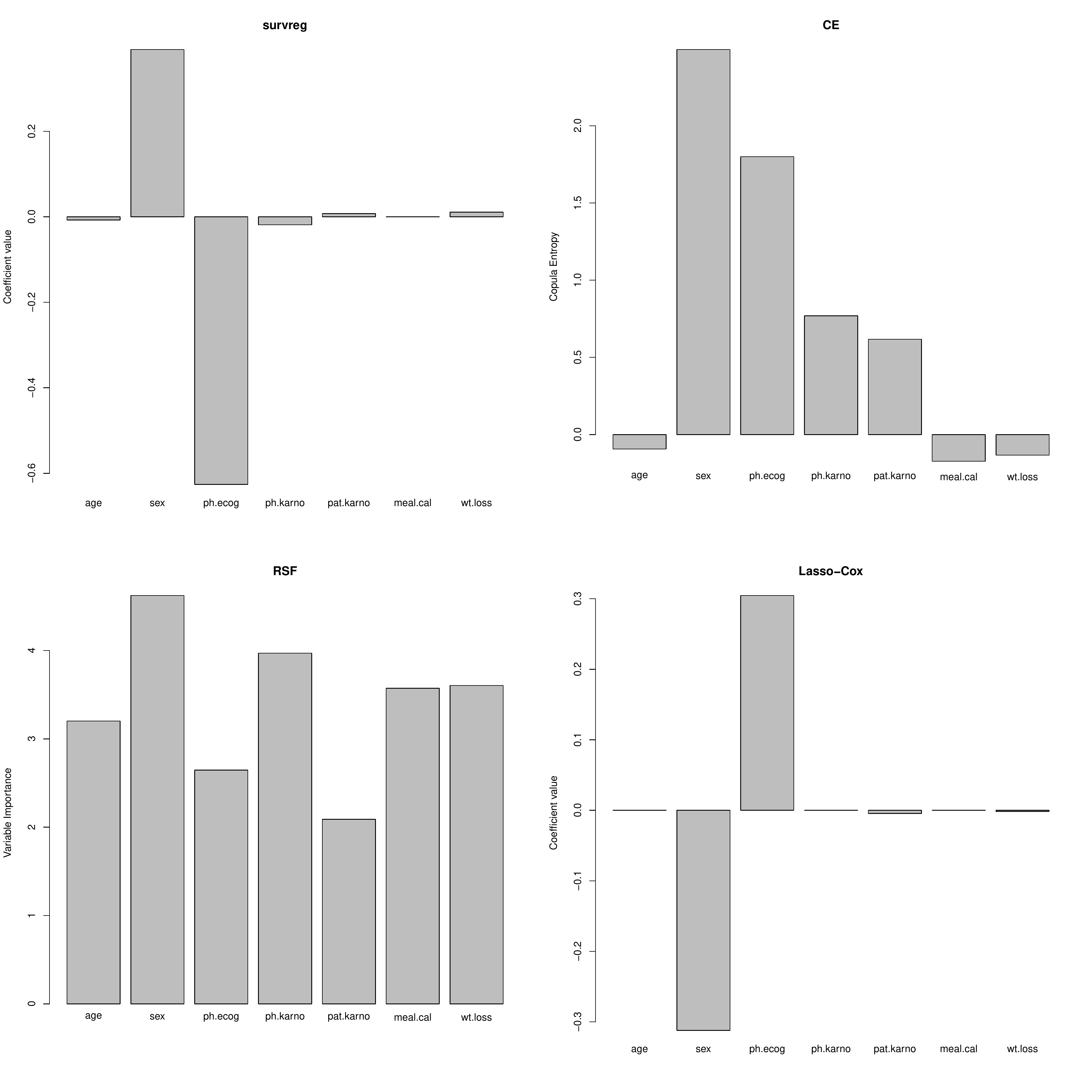}
	\caption{Variable selection with the four models on the cancer data.}
	\label{fig:cancervar}
\end{figure}

The variable importance of the selected variables of the four models on the cancer data are shown in Figure \ref{fig:cancervar} and the selected variables are listed in Table \ref{tab:cancervar}. We can learn that our method selects 4 out of 7 variables (sex, ECOG performance score and Karnofsky performance score by physicians, and Karnofsky performance score by patients). In the results of survival regression model and Lasso-Cox model, both sex and ECOG performance score by physicians has much larger importance value than other variables. RSF selects 3 variables out.

\begin{figure}
	\centering
	\includegraphics[width=\linewidth]{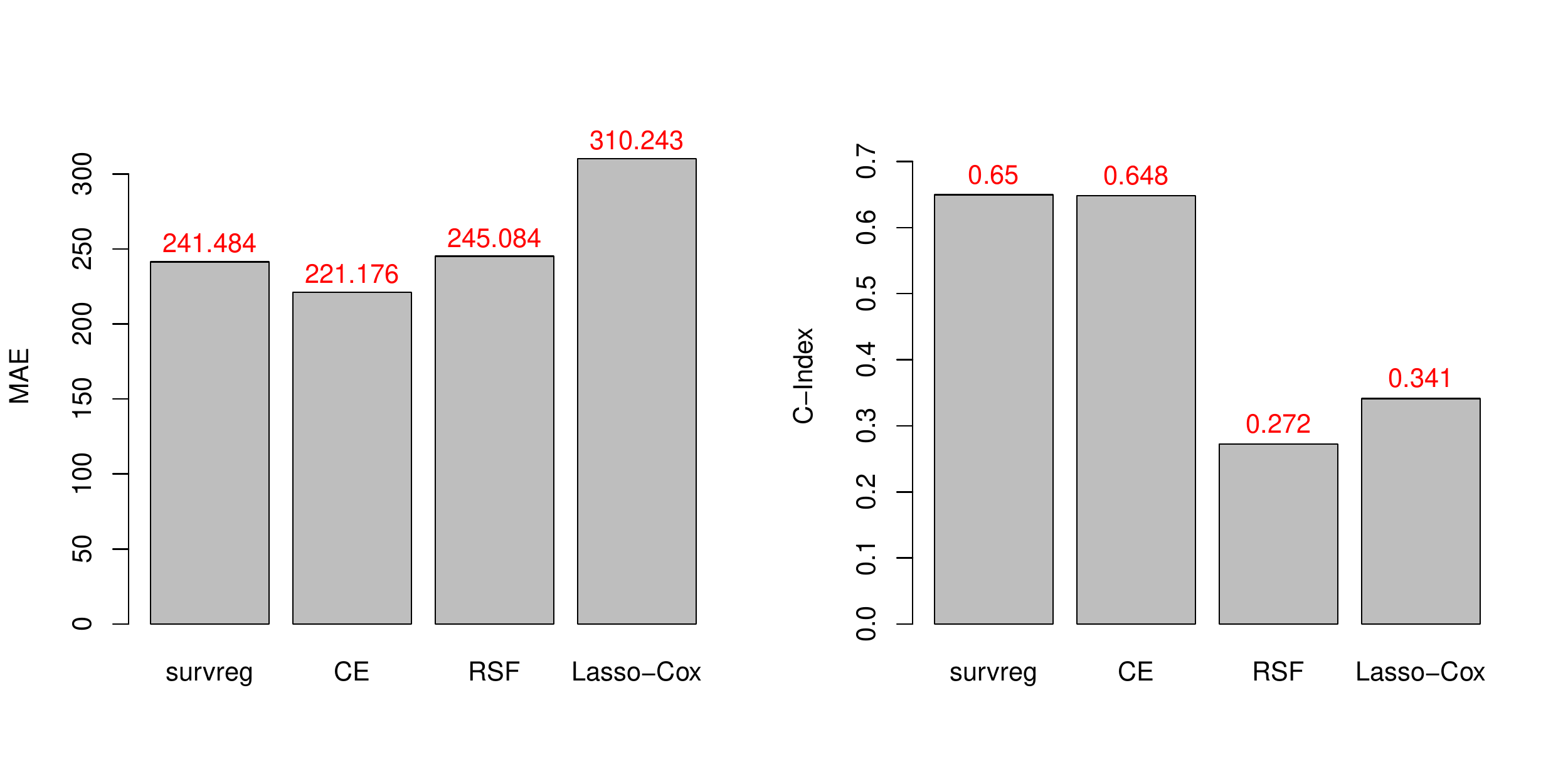}
	\caption{Prediction performance of the four models on the cancer data.}
	\label{fig:cancermeasure}
\end{figure}

The prediction performance of the four model on the cancer data in terms of MAE and C-Index are shown in Figure \ref{fig:cancermeasure}. It can be learned from it that our method presents better results than survival regression model with smaller MAE score and comparable C-Index score while the other two models present worse results than these two models with larger MAE scores and much smaller C-Index scores.

\begin{figure}
	\centering
	\includegraphics[width=\linewidth]{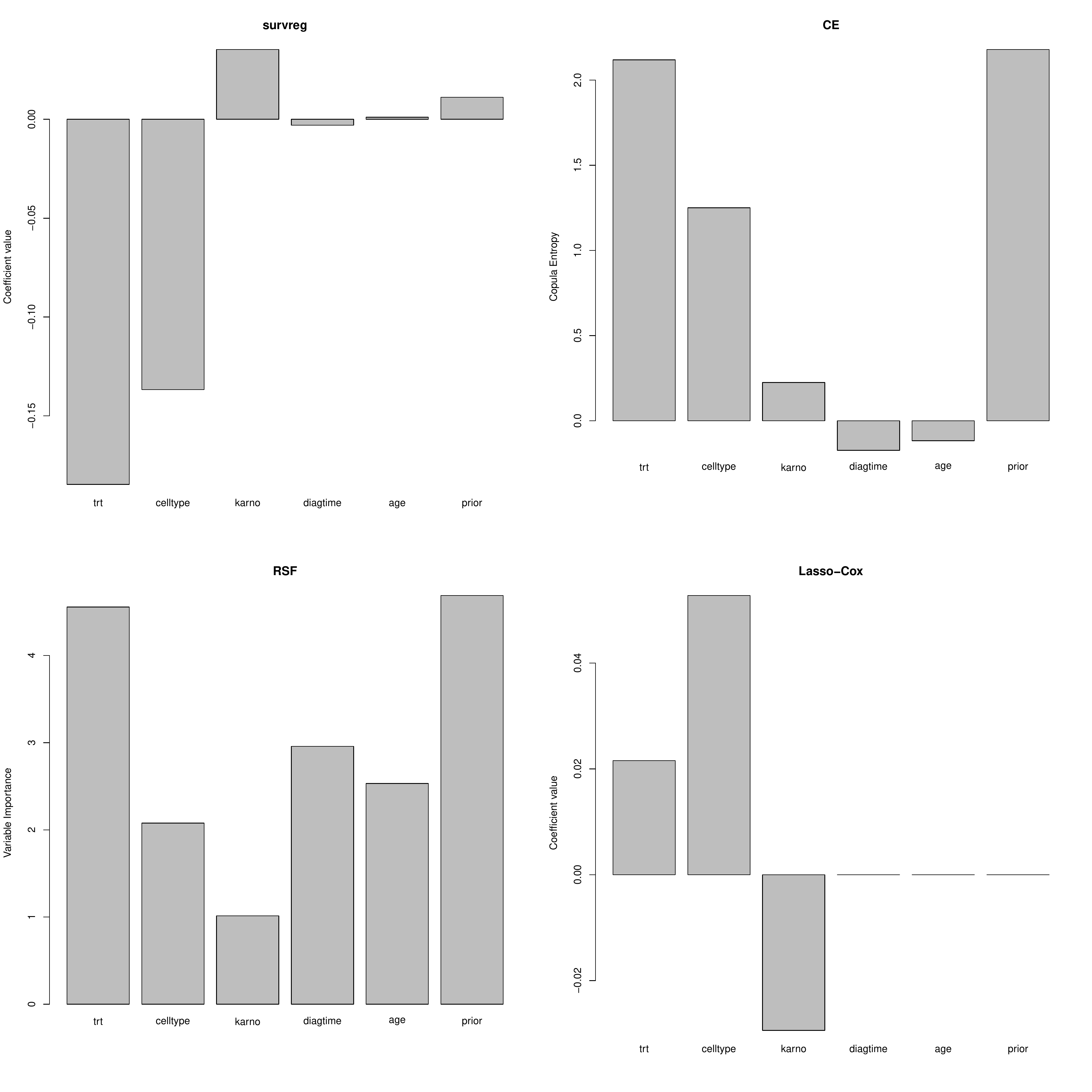}
	\caption{Variable selection with the four models on the veteran data.}
	\label{fig:veteranvar}
\end{figure}

The variable importance of the selected variables of the four models on the veteran data are shown in Figure \ref{fig:veteranvar} and the selected variables are listed in Table \ref{tab:veteranvar}. Our method selects 4 out of 6 variables (test type, cell type, Karnofsky performance score, and the indicator of prior treatment). The survival regression model selects 5 variables (a weight loss more than our method) according to the coefficients of the covariates of the model. The Lasso-Cox model selects 3 variables (test type, cell type, and Karnofsky performance score). 3 variables are selected with RSF: age, cell type, and Karnofsky performance score.

\begin{figure}
	\centering
	\includegraphics[width=\linewidth]{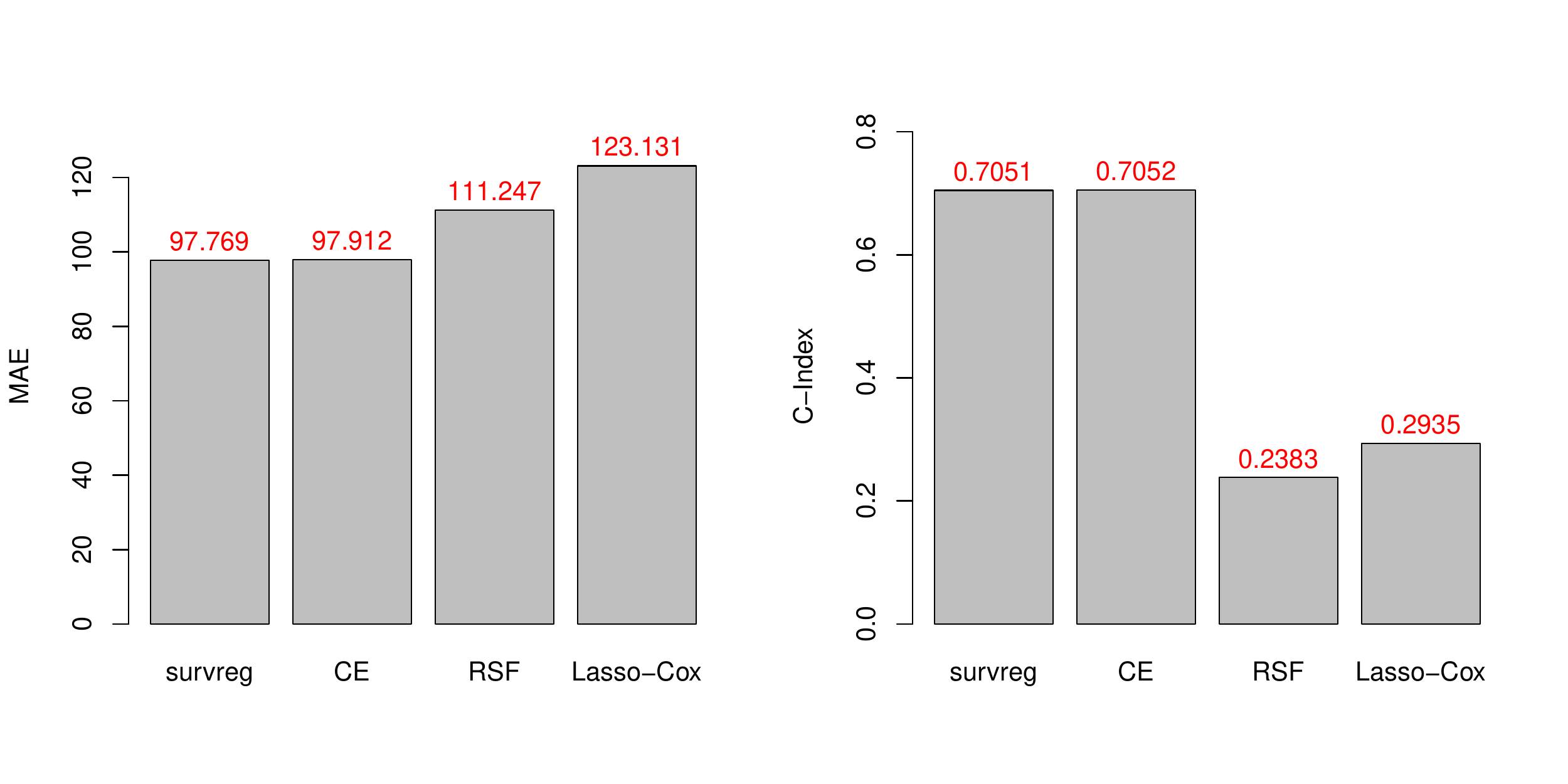}
	\caption{Prediction performance of the four models on the veteran data.}
	\label{fig:veteranmeasure}
\end{figure}

The prediction performance of the four model on the veteran data in terms of MAE and C-Index are shown in Figure \ref{fig:veteranmeasure}. It is similar to that of the cancer data. Our method presents comparable MAE score and larger C-Index score than the survival regression model while RSF and Lasso-Cox only present much worse results than these two models.

\begin{table}
	\centering
	\caption{Selected variables with the four models on the cancer data.}
	\begin{tabular}{l|c}
		\toprule
		Model&Selected variables\\
		\midrule
		survreg&age, sex, ph.ecog, ph.karno, pat,karno, wt.loss\\
		CE&sex, ph.ecog, ph.karno, pat.ecog\\
		RSF&age, ph.ecog, pat.karno\\
		Lasso-Cox&sex, ph.ecog, pat.karno, wt.loss\\
		\bottomrule
	\end{tabular}
	\label{tab:cancervar}
\end{table}

\begin{table}
	\centering
	\caption{Selected variables with the four models on the veteran data.}
	\begin{tabular}{l|c}
		\toprule
		Model&Selected variables\\
		\midrule
		survreg&trt, celltype, karno, prior, diagtime\\
		CE&trt, celltype, karno, prior\\
		RSF&age, celltype, karno\\
		Lasso-Cox&trt, celltype, karno\\
		\bottomrule
	\end{tabular}
	\label{tab:veteranvar}
\end{table}

\section{Discussion}
\label{sec:discussion}

In simulation experiment, we simulated a sample of survival model with 5 covariates. We attach the covariates with different importance by different coefficient values. As shown in Figure \ref{fig:sim1}, our method can estimate the right importance of covariates with estimated CE values with reference to the coefficients $\beta$ of the covariates. Moreover, there is a little difference between the estimated results of with and without censoring information. These results show the effectiveness of our method for VS on survival analysis.

For the cancer data, the four trained models presents different selected variables, as listed in Table \ref{tab:cancervar}. Sex is a well-known prognostic factor of lung cancer, that are selected by all the four models. Performance scores are the common factors for lung cancer \cite{Brundage2002} and are widely considered as more important than the other factors \cite{Buccheri1994}. In the experiment our method selected all the three performance factors out, which show its effectiveness for clinical use, while RSF and Lasso-Cox select 2 factors (ECOG performance score and Karnofsky performance score by patients). Weight loss is considered as an indicator for risk factor. Here only RSF select it out.

The veteran data is also collected from patients with lung cancer but contains different variables with the cancer data. This time our method selected 4 variables (test type, cell type, Karnofsky score and prior treatment) out. The cell type and Karnofsky score factors are also selected by the other two models. Our method also selected the factor for priori treatment, which can be explained as related to another important prognostic factor -- cancer stage \cite{Brundage2002,Buccheri1994}. 

It can learned from Figure \ref{fig:cancermeasure} and \ref{fig:veteranmeasure} that for both data, the model built with the selected variables with our method presents better prediction performance than RSF and Lasso-Cox did. Compared with the survival regression model built with all the variables, our method presents competitive performance with better MAE on the cancer data and better C-Index on the veteran data. This result is achieved with fewer variables than the survival regression model with full variables. This implies that our method can select more interpretable variables without sacrificing predictability.

\section{Conclusions}
\label{sec:con}
In this paper we propose to apply CE-based VS method to survival analysis. The idea is to measure the correlation between variables and time-to-event with CE and then select the variables associated with smaller CE values. CE is a model-free measure for statistical independence and has a non-parametric estimation method. Therefore, the proposed method can be applied to survival analysis without any assumptions. It is also a tuning-free method compared with the method, such as Lasso-Cox that need tuning hyperparameter for variable selection. We did simulation experiment to validate the effectiveness of the proposed method. Experiments with two real data were also performed to compare the proposed method with two other related methods: RSF and Lasso-Cox. Experimental results showed that our method can select interpretable variables out without sacrificing prediction performance and presented better results than these two related methods in terms of both interpretability and predictability.

\bibliographystyle{unsrt}
\bibliography{surv}

\end{document}